\documentclass[aps,pre,showpacs,twocolumn,superscriptaddress,floatfix]{revtex4}
\usepackage{graphicx,color,epsfig} 
\usepackage{color}
\usepackage[T1]{fontenc}
\usepackage[utf8x]{inputenc}
\usepackage{lipsum}
\usepackage{amsmath}
\usepackage[title]{appendix}
\usepackage{amssymb}

\begin{document}

%\title{An agent-based model in the context of the Mafia game to describe the dissipation of disinformation}
\title{Cumulative suspicion and absorption dynamics in an agent-based Mafia game
%A Mafia Game–Based Model for the Propagation and Dissipation of Disinformation
}

\author{Eduardo Velasco Stock}
\email{eduardo.stock@ufrgs.br}

\author{Roberto da Silva}
\email{rdasilva@if.ufrgs.br}

\affiliation{Instituto de Física, Universidade Federal do Rio Grande do Sul, Caixa Postal 15051, 91501-970 Porto Alegre RS, Brazil}

\date{\today}

\begin{abstract}

The Mafia game, also known as Werewolf, describes a competition between a coordinated minority whose identities are concealed and a larger good faction composed primarily of uninformed civilians attempting to identify and eliminate them. We introduce an agent-based formulation in which the daytime decision is not represented by uniform random voting. Instead, randomly paired agents modify player-specific public suspicion scores according to their roles, and one surviving player is subsequently eliminated with probability proportional to their accumulated suspicion. These scores persist throughout the game, producing a history-dependent stochastic process with two competing absorbing outcomes: elimination of the mafia or numerical parity between the mafia and the good faction. We investigate the effects of population size, initial mafia size, and the presence of perfectly informed detectives through extensive Monte Carlo simulations. In the absence of detectives, a random-execution approximation predicts that, in the dilute-mafia and early-time regime, the cumulative probability of mafia extinction behaves as $F(\tau)\sim (\tau/N)^{N_m}$, in agreement with simulations as $N_m/N$ decreases. The winning-probability curves also exhibit an empirical finite-size crossover characterized by a population scale $N_c$. Rescaling the population by $N_c$ approximately collapses the curves obtained for different initial mafia populations. Perfectly informed detectives substantially shorten mafia-survival times and introduce an additional dependence on population composition for which the same one-parameter rescaling is insufficient. The model provides a minimal connection between microscopic histories of accusation and the macroscopic absorption statistics of hidden-role games.

\end{abstract}

\maketitle

\section{Introduction}

The application of statistical-physics methods to social systems has produced a broad class of minimal models to investigate how microscopic interactions give rise to collective behavior \cite{Castellano2009,Galam2012}. Examples include stochastic models of spatial conflict and opinion exchange \cite{Clifford1973,Holley1975}, cultural dissemination \cite{Axelrod1997}, rumor propagation \cite{Daley1965,Maki1973}, minority influence \cite{Galam2002}, and majority-rule dynamics \cite{Krapivsky2003}. Although these formulations simplify human behavior, they offer controlled environments to isolate the effects of population composition, information asymmetry, interaction rules, and stochastic fluctuations.

Voter-model studies with zealots provide a closely related physical example: even a small group of permanently committed agents can change the relaxation and long-time behavior of collective opinion dynamics \cite{Mobilia2003}.

A particularly compelling domain involves systems where a coordinated minority uses privileged information to influence a less informed majority. Related phenomena appear in rumors, misinformation, and political communication, where persistent narratives depend on repeated exposure, group identity, and strategic coordination \cite{Lazer2018,Vosoughi2018}. While real-world dynamics are far more complex, they motivate a fundamental question: under what conditions can a majority identify and eliminate an informed, manipulating minority? Related game-theoretic formulations have examined the collective consequences of asymmetric access to information. In the boxers game, for example, informed agents transmit truthful or deceptive information to uninformed players through pairwise interactions \cite{DASILVA2018283}.

The Mafia party game, also known as Werewolf party game, offers a clear framework for studying such conflicts. The system comprises an informed minority—the mafia—and an uninformed majority of ordinary citizens. Gameplay alternates between a private phase (the mafia eliminates a member of the good faction) and a public phase (where the surviving population votes to eliminate a suspected player). The game terminates in competing absorbing states: citizens win if all mafia members are removed, whereas the mafia wins when its surviving population becomes at least as large as the good faction.

From a statistical-physics perspective, these termination rules place the game within the broad framework of stochastic dynamics with absorbing states, for which fluctuations, finite size, and the structure of the absorbing configurations can control the macroscopic behavior \cite{Hinrichsen2000,Odor2004,Henkel2008,Henkel2010}.

Despite these features, the mathematical literature on the Mafia game remains limited. Braverman et al. \cite{Braverman2008} established that, without detectives, the mafia and citizens have comparable winning probabilities when the initial mafia size scales with the square root of the population. They also showed that the introduction of informed detectives qualitatively changes this scaling. Migdał \cite{Migdal2010} mapped a simplified version to a discrete pure-death process, deriving exact analytical expressions that highlighted finite-size and parity effects. Other studies have explored optimal strategies under alternative rules \cite{Wang2024}. Concurrently, Mafia and Werewolf have been adopted as testbeds for artificial agents, strategic communication, and large-language-model reasoning \cite{Eger2018,Xu2023,Bailis2024}. Other studies have used human gameplay to investigate linguistic and multimodal indicators of deception and persuasion, including the use of language models to identify concealed roles \cite{Yoo2024,ibraheem2022,Demyanov2015,lai2022}. However, these computational approaches focus primarily on individual-level inference, communication, and gameplay rather than on minimal population-level stochastic dynamics.

Existing mathematical models typically simplify the public phase using uniformly random voting, aggregate transition rates, or fixed strategies, while AI-driven models focus on detailed dialogue and reasoning. An intermediate agent-based framework—where simple pairwise interactions generate an evolving collective tendency to eliminate specific players—remains underexplored.

In this work, we introduce an agent-based model of the Mafia game in which public eliminations are mediated by cumulative, player-specific suspicion. Pairwise stochastic interactions among mafia members, civilians, and detectives continuously update individual suspicion scores. Consequently, two realizations with identical faction populations may have different public-elimination probabilities because their accumulated suspicion configurations differ.

Using Monte Carlo simulations, we examine how this history-dependent mechanism drives macroscopic dynamics, focusing on population trajectories, mafia extinction times, and winning probabilities. Without detectives, the early-time mafia-extinction statistics approach the prediction of an analytically tractable random-elimination benchmark as the initial mafia fraction decreases. Furthermore, the winning-probability curves obtained for different initial mafia populations can be approximately collapsed through a one-parameter finite-size rescaling. Introducing detectives accelerates mafia extinction and prevents the same one-parameter rescaling from collapsing the winning-probability curves, indicating that the internal composition of the good faction introduces an additional control variable.

Rather than capturing every psychological or linguistic nuance of real gameplay, our model provides a minimal stochastic framework linking microscopic histories of accusation to macroscopic absorption statistics. Broadly, it serves as an abstract representation of collective decision-making under concealed affiliations and imperfect inference.

The remainder of this paper is organized as follows. Section II defines the microscopic dynamics, interaction rules, and observables. Section III derives the analytical benchmark for early-time extinction statistics. Section IV reports numerical results for temporal evolution, extinction times, and winning probabilities. Section V concludes with a discussion, conclusions, and future directions.

\section{\label{sec:model}Agent-based Mafia model}
%======================================================================

We consider a well-mixed population of $N(t)$ surviving agents at
round $t$. Each agent belongs to one of three subpopulations: mafia
members, detectives, or civilians. Their respective numbers are
denoted by $N_m(t)$, $N_d(t)$, and $N_c(t)$, so that
\begin{equation}
    N(t)=N_m(t)+N_d(t)+N_c(t).
\end{equation}
Detectives and civilians form the good faction, whose population is
\begin{equation}
    N_g(t)=N_d(t)+N_c(t).
\end{equation}
The initial population and subpopulation sizes are denoted by
$N_0=N(0)$ and $N_{x,0}=N_x(0)$, with $x=m,d,c,g$.

Each surviving agent $i$ is also associated with a non-negative
public-suspicion score $a_i(t)$. Initially,
\begin{equation}
    a_i(0)=0
\end{equation}
for every agent. The scores are cumulative: unless an agent is
eliminated, the suspicion acquired during one round is retained in all
subsequent rounds. Thus, the complete microscopic state of the system
is determined not only by the surviving numbers of each role, but
also by the configuration of suspicion scores,
\begin{equation}
    \mathbf{a}(t)=\{a_i(t):i\in\mathcal{L}(t)\},
\end{equation}
where $\mathcal{L}(t)$ denotes the set of surviving agents.

The game has two absorbing outcomes. The good faction wins when
\begin{equation}
    N_m(t)=0,
\end{equation}
whereas the mafia wins when it reaches numerical parity with the good
faction,
\begin{equation}
    N_m(t)\geq N_g(t).
\end{equation}
The absorbing conditions are checked after each elimination. In
particular, if parity is reached during the night phase, the game ends
before the corresponding day phase.

Each nonabsorbed round consists of a night phase followed, when
applicable, by a day phase.

%----------------------------------------------------------------------
\subsection{Night phase}
%----------------------------------------------------------------------

During the night, one surviving member of the good faction is selected
uniformly at random and eliminated by the mafia. Therefore, for
$i\in\mathcal{G}(t)$,
\begin{equation}
    \Pr(i\text{ is eliminated at night})
    =
    \frac{1}{N_g(t)},
\end{equation}
where $\mathcal{G}(t)$ is the set of surviving detectives and
civilians. Mafia members cannot be eliminated during this stage.

After the night elimination, the mafia-win condition is tested. If
the surviving mafia population is at least as large as the surviving
good population, the game is over. Otherwise, the
surviving agents proceed to the day phase.

%----------------------------------------------------------------------
\subsection{Day phase and suspicion dynamics}
%----------------------------------------------------------------------

At the beginning of the day phase, the surviving agents are randomly
shuffled and arranged into disjoint pairs. Each pair produces one unit
of public suspicion, distributed according to the roles of the
interacting agents. All target selections described below are uniform
over the corresponding eligible set.

Let $\mathcal{M}$, $\mathcal{D}$, and $\mathcal{C}$ denote the sets of
surviving mafia members, detectives, and civilians during the current
day phase, and let $\mathcal{G}=\mathcal{D}\cup\mathcal{C}$.

The suspicion-update rules are as follows:

\begin{itemize}

    \item \textbf{Mafia--mafia interaction.}
%    The two mafia members direct suspicion toward a randomly selected
%    member $k$ of the good faction:
    The two mafia members direct suspicion toward a member $k$ of the good 
    faction, selected uniformly at random:
    \begin{equation}
        k\sim\mathcal{U}(\mathcal{G}),
        \qquad
        a_k\leftarrow a_k+1.
    \end{equation}

    \item \textbf{Mafia--civilian interaction.}
    The mafia member redirects the civilian's suspicion toward a
    randomly selected good agent other than the civilian participating
    in the pair. If $j$ denotes the paired civilian,
    \begin{equation}
        k\sim\mathcal{U}(\mathcal{G}\setminus\{j\}),
        \qquad
        a_k\leftarrow a_k+1.
    \end{equation}

    \item \textbf{Mafia--detective interaction.}
    The detective correctly recognizes the mafia member, whereas the
    mafia member attempts to redirect suspicion toward the detective.
    The two agents consequently receive equal increments:
    \begin{equation}
        a_m\leftarrow a_m+\frac{1}{2},
        \qquad
        a_d\leftarrow a_d+\frac{1}{2},
    \end{equation}
    where $m$ and $d$ denote the paired mafia member and detective,
    respectively.

    \item \textbf{Civilian--civilian interaction.}
    Because neither civilian possesses privileged information, the
    pair directs suspicion toward a randomly selected surviving agent
    other than the two participants:
    \begin{equation}
        k\sim\mathcal{U}
        \bigl(\mathcal{L}\setminus\{i,j\}\bigr),
        \qquad
        a_k\leftarrow a_k+1.
    \end{equation}

    \item \textbf{Detective--civilian or detective--detective
    interaction.}
    The informative contribution of the detective or detectives is
    directed toward a genuine surviving mafia member:
    \begin{equation}
        k\sim\mathcal{U}(\mathcal{M}),
        \qquad
        a_k\leftarrow a_k+1.
    \end{equation}

\end{itemize}

The detective dynamics is implemented at a coarse-grained level.
Detectives are treated as perfectly informed contributors to the
public-suspicion process: whenever their interaction is informative,
the corresponding accusation is directed toward an actual surviving
mafia member. The model does not explicitly store private,
agent-specific histories of previously investigated identities.

If the number of surviving agents entering the day phase is odd, one
agent remains outside the pair matching. In the implementation, this
unpaired agent produces a half-weight contribution by directing
suspicion toward a uniformly selected other survivor,
\begin{equation}
    k\sim\mathcal{U}
    \bigl(\mathcal{L}\setminus\{i\}\bigr),
    \qquad
    a_k\leftarrow a_k+\frac{1}{2}.
\end{equation}
This convention preserves the same average contribution of one
half-unit of suspicion per active agent during the day phase.

After all interactions have occurred, let
\begin{equation}
    A(t)=\sum_{j\in\mathcal{L}(t)}a_j(t)
\end{equation}
be the total suspicion carried by the surviving population. The public
elimination is represented by a single stochastic collective
selection. Agent $i$ is eliminated with probability
\begin{equation}
    p_i(t)
    =
    \frac{a_i(t)}{A(t)}.
    \label{eq:weighted_execution}
\end{equation}
If $A(t)=0$, the eliminated agent is instead selected uniformly from
the surviving population. Once an agent is eliminated, that agent and
the associated suspicion score are removed from the active state.

The good-win and mafia-win conditions are tested again after the
public elimination. If neither condition is satisfied, the game
continues to the following round.

Equation~\eqref{eq:weighted_execution} implies that the faction
populations alone do not determine the subsequent transition
probabilities. Two realizations having identical values of
$N_m(t)$, $N_d(t)$, and $N_c(t)$ may have different public-elimination
probabilities because their accumulated suspicion configurations are
different.

%----------------------------------------------------------------------
\subsection{Observables and simulation protocol}
%----------------------------------------------------------------------

To compare the temporal evolution of systems with different initial
sizes, we define the normalized abundance of subpopulation $x$ as
\begin{equation}
    \rho_x(t)
    =
    \frac{N_x(t)}{N_0},
    \qquad
    x=m,d,c,g,
    \label{eq:normalized_abundance}
\end{equation}
with
\begin{equation}
    \rho_g(t)=\rho_d(t)+\rho_c(t).
\end{equation}
The quantities in Eq.~\eqref{eq:normalized_abundance} are normalized
with respect to the initial population and should not be confused
with the instantaneous faction fractions $N_x(t)/N(t)$.

The absorption time is therefore a first-passage observable: it records the first round in which a trajectory reaches either terminal boundary, connecting the analysis to the standard theory of first-passage processes \cite{Redner2001}.

For each realization $r$ of the game, we record the absorbing outcome and its
absorption time $T_r$. We introduce the indicator
\begin{equation}
    I_r=
    \begin{cases}
        1, & \text{if the good faction wins realization }r,\\
        0, & \text{if the mafia wins realization }r.
    \end{cases}
\end{equation}
For fixed initial conditions, the good-faction winning probability is
estimated as
\begin{equation}
    \widehat{\Pi}_g
    \left(N_0;N_{m,0},N_{d,0}\right)
    =
    \frac{1}{N_{\mathrm{run}}}
    \sum_{r=1}^{N_{\mathrm{run}}}I_r,
    \label{eq:good_win_probability}
\end{equation}
and
\begin{equation}
    \widehat{\Pi}_m=1-\widehat{\Pi}_g.
\end{equation}

% For realizations ending in a good-faction victory, the absorption time
% is also the mafia-extinction time, denoted by $T_m$. Because mafia
% victory is a competing absorbing outcome, one may distinguish the
% cumulative incidence of mafia extinction,
% \begin{equation}
%     \widehat{F}_g(\tau)
%     =
%     \frac{1}{N_{\mathrm{run}}}
%     \sum_{r=1}^{N_{\mathrm{run}}}
%     I_r\,\mathbb{I}(T_r\leq\tau),
%     \label{eq:cumulative_incidence}
% \end{equation}
% from the extinction-time distribution conditioned on good-faction
% victory,
% \begin{equation}
%     \widehat{F}_{m|g}(\tau)
%     =
%     \frac{
%     \displaystyle
%     \sum_{r=1}^{N_{\mathrm{run}}}
%     I_r\,\mathbb{I}(T_r\leq\tau)
%     }{
%     \displaystyle
%     \sum_{r=1}^{N_{\mathrm{run}}}I_r
%     }.
%     \label{eq:conditional_extinction_cdf}
% \end{equation}
% Here, $\mathbb{I}(\cdot)$ is the indicator function. The
% extinction-time distributions reported below are conditioned on
% good-faction victory unless stated otherwise.

All results are obtained from independent Monte Carlo realizations.
Unless indicated otherwise, averages and estimated probabilities are
computed from $N_{\mathrm{run}}=2\times10^5$ realizations.

\section{\label{sec:mean_field}Mean--Field analysis}

Here, we present analytical results within a mean-field approximation. Although our derivation was developed independently of the work of P. Migdał \cite{Migdal2010}, we subsequently identified substantial overlap with his findings. Therefore, the analysis presented below should be regarded as an alternative formulation of the mean-field regime of the game rather than as an entirely original result.

After $k$ rounds, the total number of individuals in the population is
\begin{equation}
N_k=N-2k,
\end{equation}
since two individuals are removed at each round. Let $N_m$ be the initial
number of mafia members. We denote by $X_k$ the random variable representing the
number of mobsters that have been eliminated after $k$ rounds. Thus,
\begin{equation}
P_k(j)=\Pr(X_k=j)
\end{equation}
is the probability that exactly $j$ mobsters have died up to round $k$.
The evolution of $P_k(j)$ is governed by the recurrence relation
\begin{equation}
P_{k+1}(j)
=
w_k(j-1\rightarrow j)P_k(j-1)
+
w_k(j\rightarrow j)P_k(j).
\label{Eq_recurrence}
\end{equation}

The transition probability from $j-1$ to $j$ corresponds to the death of one
additional mafia member. Since, in this case, the number of surviving mobsters
before the second execution is $N_m-j+1$, one has
\begin{equation}
w_k(j-1\rightarrow j)
=
\frac{N_m-j+1}{N-2k-1}.
\end{equation}
On the other hand, if $j$ mobsters have already died, the number of
surviving wrongdoers is $N_m-j$. Therefore, the probability that no additional
mobsters are eliminated in this round is
\begin{equation}
w_k(j\rightarrow j)
=
1-\frac{N_m-j}{N-2k-1}
=
\frac{N-N_m-2k+j-1}{N-2k-1}.
\end{equation}

It is important to note that the denominator is $N-2k-1$, rather than
$N-2k$, because at the $k$-th round one good citizen has already been executed. The second execution is then performed among the remaining
$N-2k-1$ individuals, and this second individual may or may not be from mafia.

Although one may attempt to solve the recurrence relation directly, an especially interesting regime is obtained when
\begin{equation}
N_m\ll N.
\end{equation}
In this limit, the probability of selecting a mafia member at each round is small, and the death times of different mobsters may be treated as approximately independent random variables.

Let us now consider one specific mobster. At round $k$, after a good citizen individual has been executed, the probability that this particular mobster is selected in the second execution is
\begin{equation}
\frac{1}{N-2k-1}.
\end{equation}
Therefore, the probability that this mafia member is not selected in that round is
\begin{equation}
1-\frac{1}{N-2k-1}.
\end{equation}
Consequently, the survival probability of this particular mobster after $t$ rounds is
\begin{equation}
q(t)
=
\prod_{k=0}^{t-1}
\left(
1-\frac{1}{N-2k-1}
\right).
\end{equation}

Taking the logarithm, we obtain
\begin{equation}
\ln q(t)
=
\sum_{k=0}^{t-1}
\ln
\left(
1-\frac{1}{N-2k-1}
\right).
\end{equation}
In the thermodynamic limit, $N\rightarrow\infty$, we use
\begin{equation}
\ln\left(1-\frac{1}{N-2k-1}\right)
\approx
-\frac{1}{N-2k-1}.
\end{equation}
Thus,
\begin{equation}
\begin{array}{ccc}
\ln q(t)
& \approx &
-\sum\limits_{k=0}^{t-1}
\frac{1}{N-2k-1}
\\[0.3cm]
& \approx &
-\int_0^t
\frac{dk}{N-2k}
\\[0.3cm]
& = &
\frac{1}{2}
\ln
\left(
1-\frac{2t}{N}
\right).
\end{array}
\end{equation}
Therefore,
\begin{equation}
q(t)
=
\sqrt{
1-\frac{2t}{N}
}.
\end{equation}

In the limit $N\gg N_m$, treating the $N_m$ mobsters as approximately independent, the number of mafia members eliminated up to time $t$ follows a binomial distribution,
\begin{equation}
P_t(j)
=
\frac{N_m!}{j!(N_m-j)!}
\left[1-q(t)\right]^j
q(t)^{N_m-j}.
\end{equation}

It is important to observe that this expression gives the probability that $j$ mobsters have died up to time $t$. Therefore, it is naturally related
to a cumulative distribution in time, since $1-q(t)$ is the probability that a given mobster has died at some time before or at $t$, rather than exactly at time $t$.

The probability that the entire mafia has been eliminated up until $\tau$, which is the quantity we calculate numerically, is then given by
\begin{equation}
\begin{array}{ccccc}
\Pr(t_{\mathrm{survival}}\leq \tau)
& = &
P_{\tau}(N_m)
& = &
\left[1-q(\tau)\right]^{N_m}
\\[0.3cm]
&  &  & = &
\left(
1-\sqrt{1-\frac{2\tau}{N}}
\right)^{N_m}.
\end{array}
\end{equation}

For early times, $\tau\ll N$, one has
\begin{equation}
\sqrt{1-\frac{2\tau}{N}}
\approx
1-\frac{\tau}{N}.
\end{equation}
Thus,
\begin{equation}
1-\sqrt{1-\frac{2\tau}{N}}
\approx
\frac{\tau}{N},
\end{equation}
and a power-law regime is expected:
\begin{equation}
\Pr(t_{\mathrm{survival}}\leq \tau)
\approx
\left(
\frac{\tau}{N}
\right)^{N_m}
\sim
\tau^{N_m}.
\label{Eq:power_law_DA}
\end{equation}

In the following section, we present our main results.

\section{\label{sec:results}Results}

\begin{figure}
    \centering
    \includegraphics[width=1.0\linewidth]{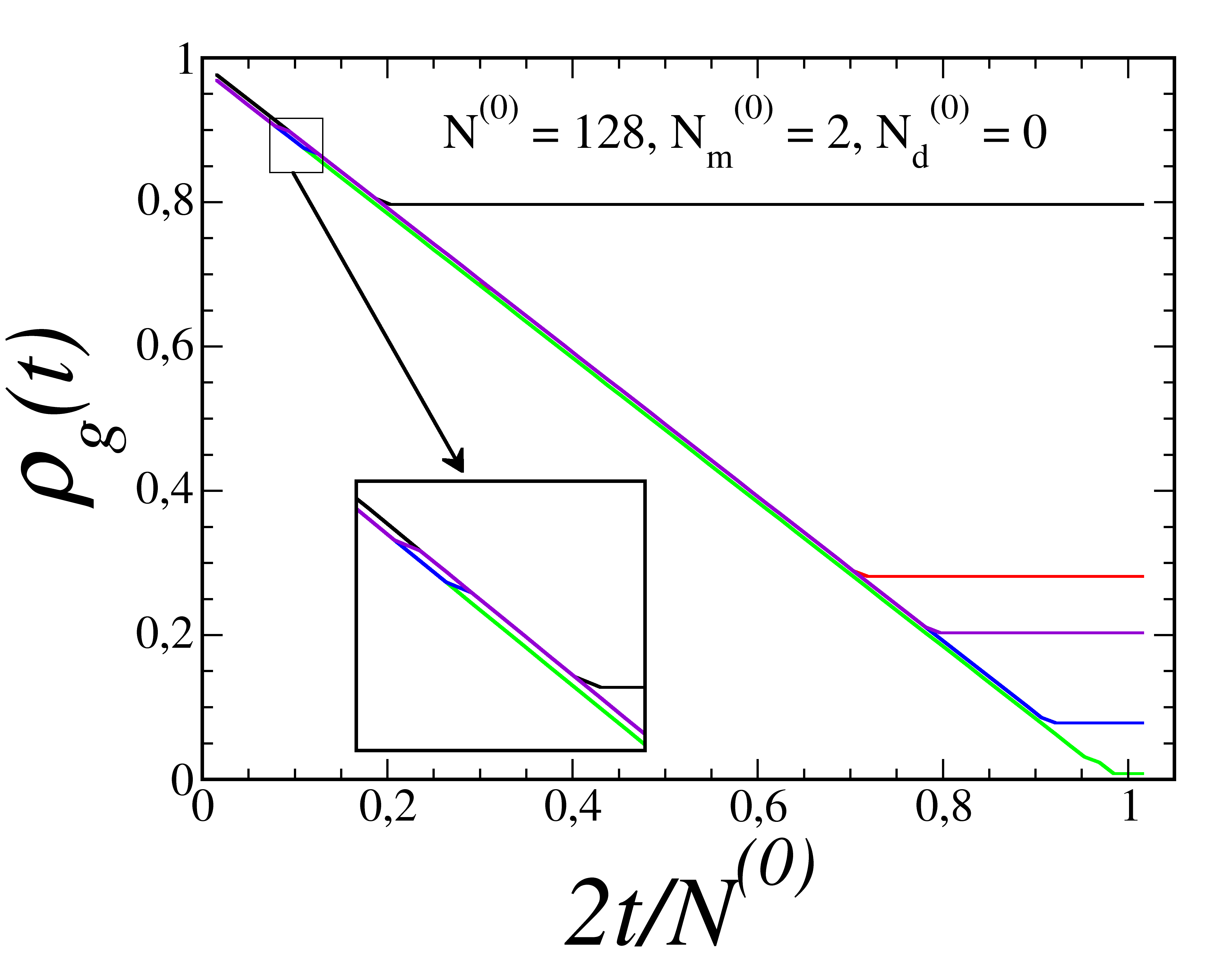}
    \caption{The typical evolution of several seeds showing the trivial linear decay of the population of good citizens, $\rho_g{(t)}$. We used $N^{(0)}=128$, $N_m^{(0)}=2$, and $N_d^{(0)}$.}
    \label{fig1_evolution}
\end{figure}

We first studied the typical evolution of the game. In Fig. \ref{fig1_evolution}, we show five different simulations with different seeds each for a system of $N^{(0)}=128$, $N_m^{(0)}=2$, and no detectives. We observe the typical trivial linear decay of the population of good citizens, which also presents, in its course, some less inclined linear segments. These discontinuities in the evolution correspond to the spurious deaths of mafia members that may happen at the public execution stage. After the entire mafia is executed, the number of good citizens stagnates at a plateau, which might happen at a random time, called the extinction time, and at a random density level.

\begin{figure}
    \centering
    \includegraphics[width=1.0\columnwidth]{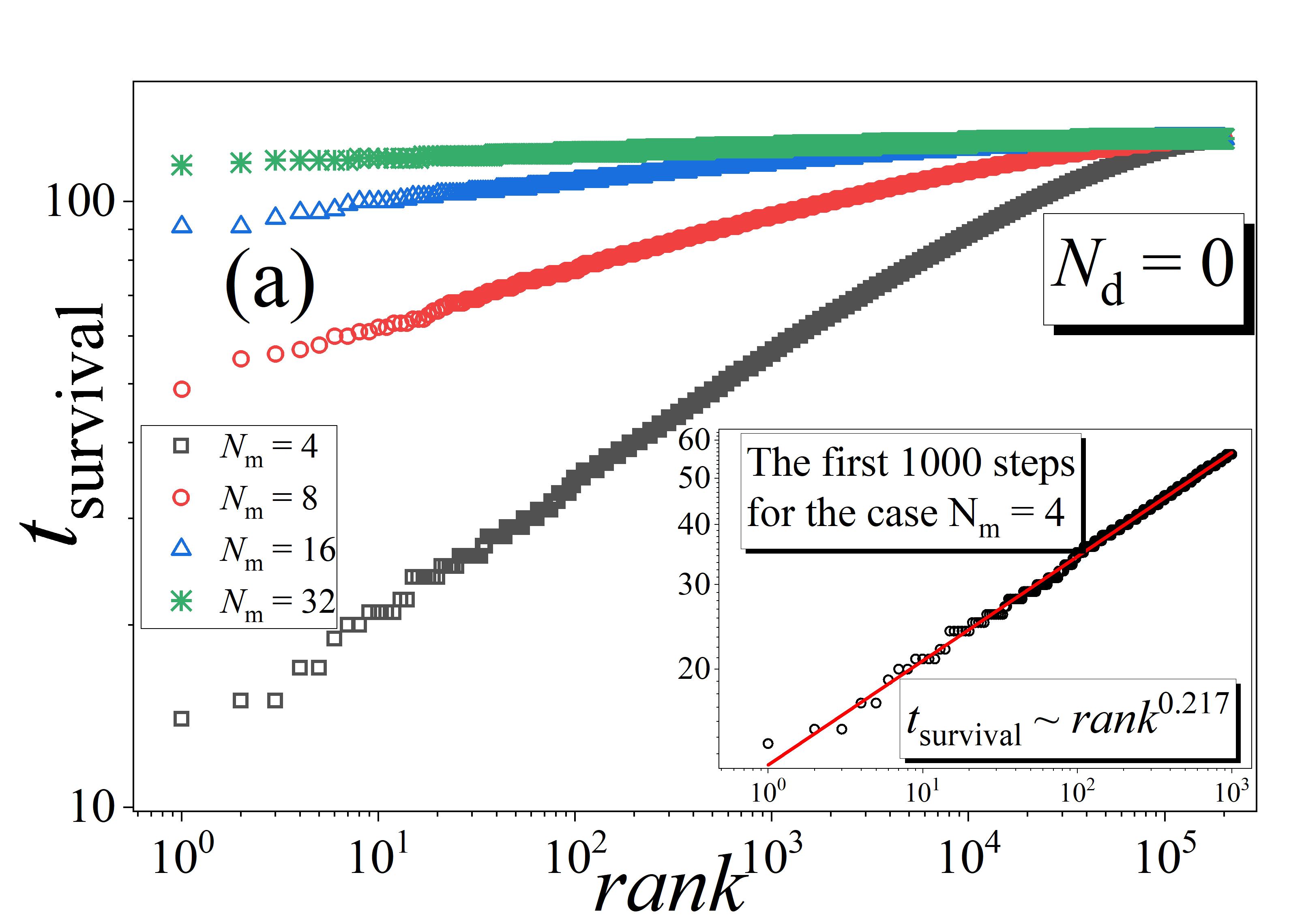}
    \includegraphics[width=1.0\columnwidth]{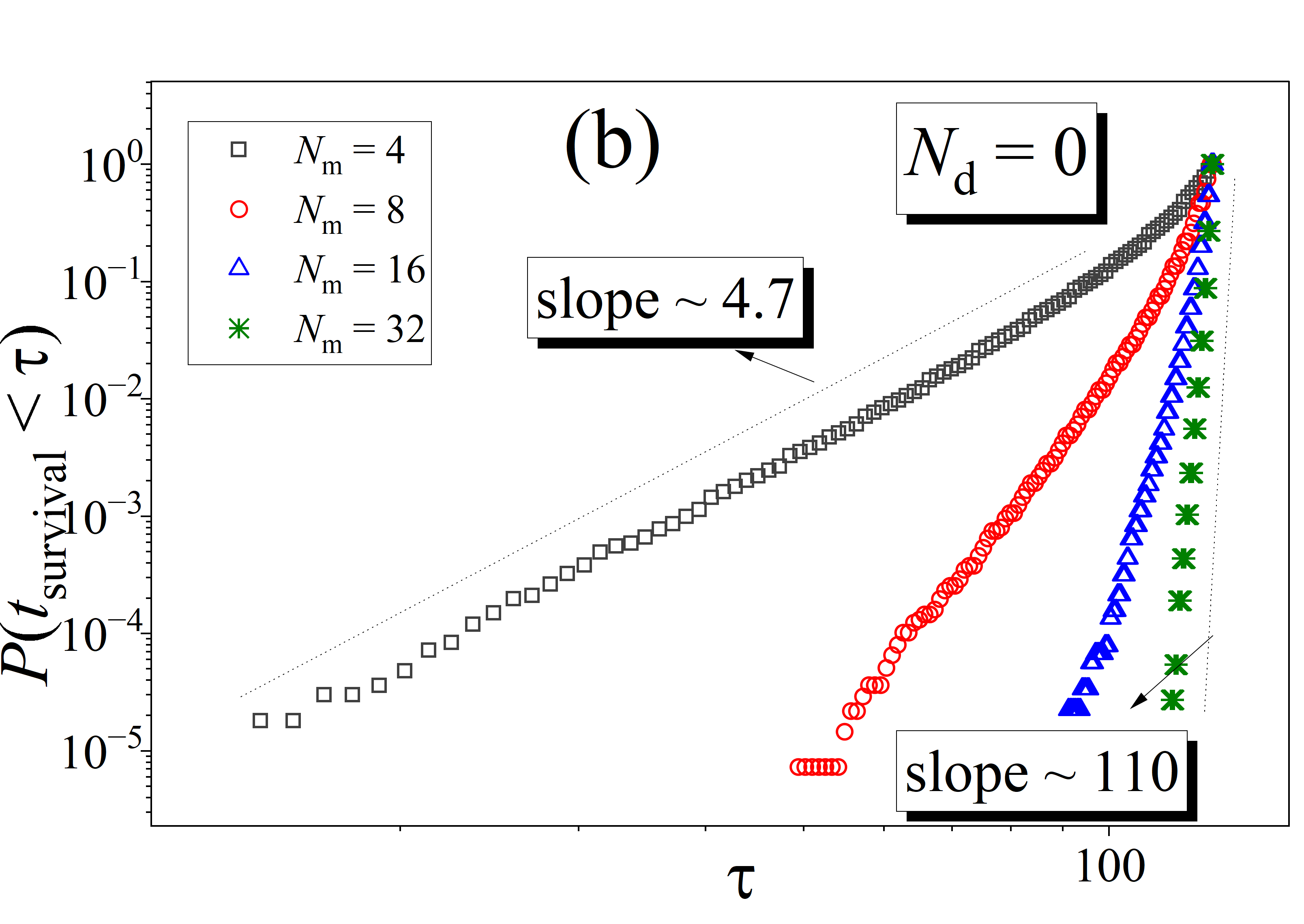}
    \includegraphics[width=1.0\columnwidth]{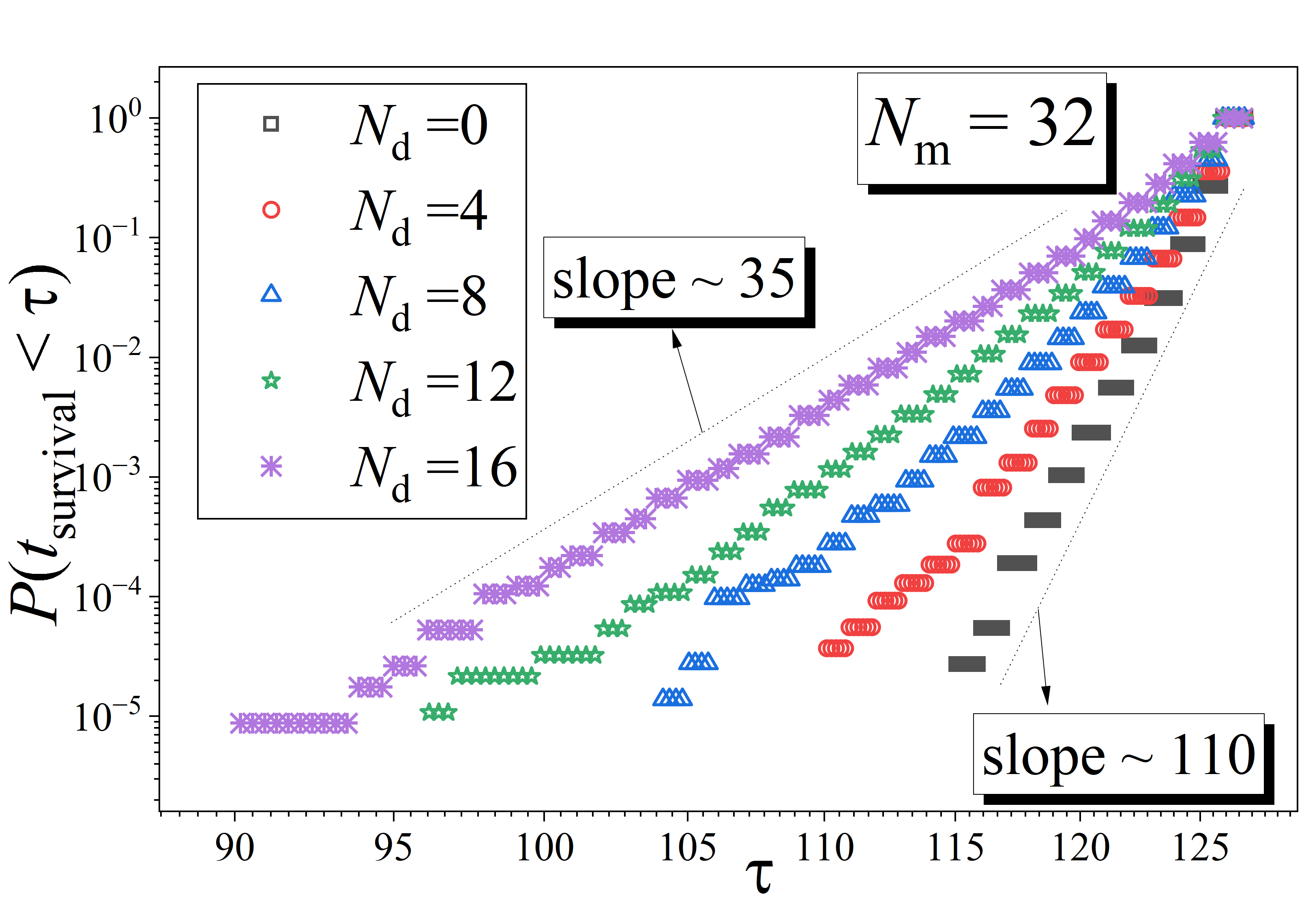}
    \caption{Survival times for the mafia. (a) Rank plot of the survival times in the absence of detectives, $n_{d}=0$; the inset highlights the power-law behavior observed for a small number of mafia members, $N_{m}=4$. (b) Cumulative distribution of survival times for $N_{d}=0$. (c) Dependence of the survival times on the number of detectives.}
    \label{fig_histogram}
\end{figure}

Thus, it is important to analyze the statistics of extermination times, which quantify how long a spreader of fake news, or simply a disinformer, can survive while sustaining misinformation in the system. We first consider the reference case corresponding to the absence of detectives, i.e., $N_{d}^{(t)}=0$. In this case, we vary the initial size of the mafia and, for each value of this parameter, we consider an ensemble of independent dynamical realizations from which the survival times are measured.

As a first analysis, all survival times are ranked in increasing order and plotted as a function of their rank, as shown in Fig.~\ref{fig_histogram}(a). The plot is presented on a log-log scale. The rank-ordered survival times reveal a clear systematic dependence on $N_{m}^{(0)}$. As $N_{m}^{(0)}$ increases, the corresponding curves are shifted upward over the entire range of ranks. This indicates that larger initial mafia size enhances the persistence of misinformation in the system, as expected. However, the rank plot allows us to quantify how this enhanced persistence occurs across different realizations.

The systematic upward displacement of the curves shows that increasing $N_{m}^{(0)}$ makes misinformation more resilient. In addition, the initial portion of the curves displays an approximate power-law behavior, suggesting a nontrivial hierarchy of survival events. In other words, although most realizations lead to relatively short survival times, a small fraction of them persists for significantly longer times. The inset of Fig.~\ref{fig_histogram}(a) illustrates this behavior for $n_{w}=4$, for which a reasonable power-law dependence is observed, approximately given by

\begin{equation}
t_{\mathrm{survival}} \sim \mathrm{rank}^{0.2}.
\end{equation}

\begin{figure}
    \centering
    \includegraphics[width=1.0\columnwidth]{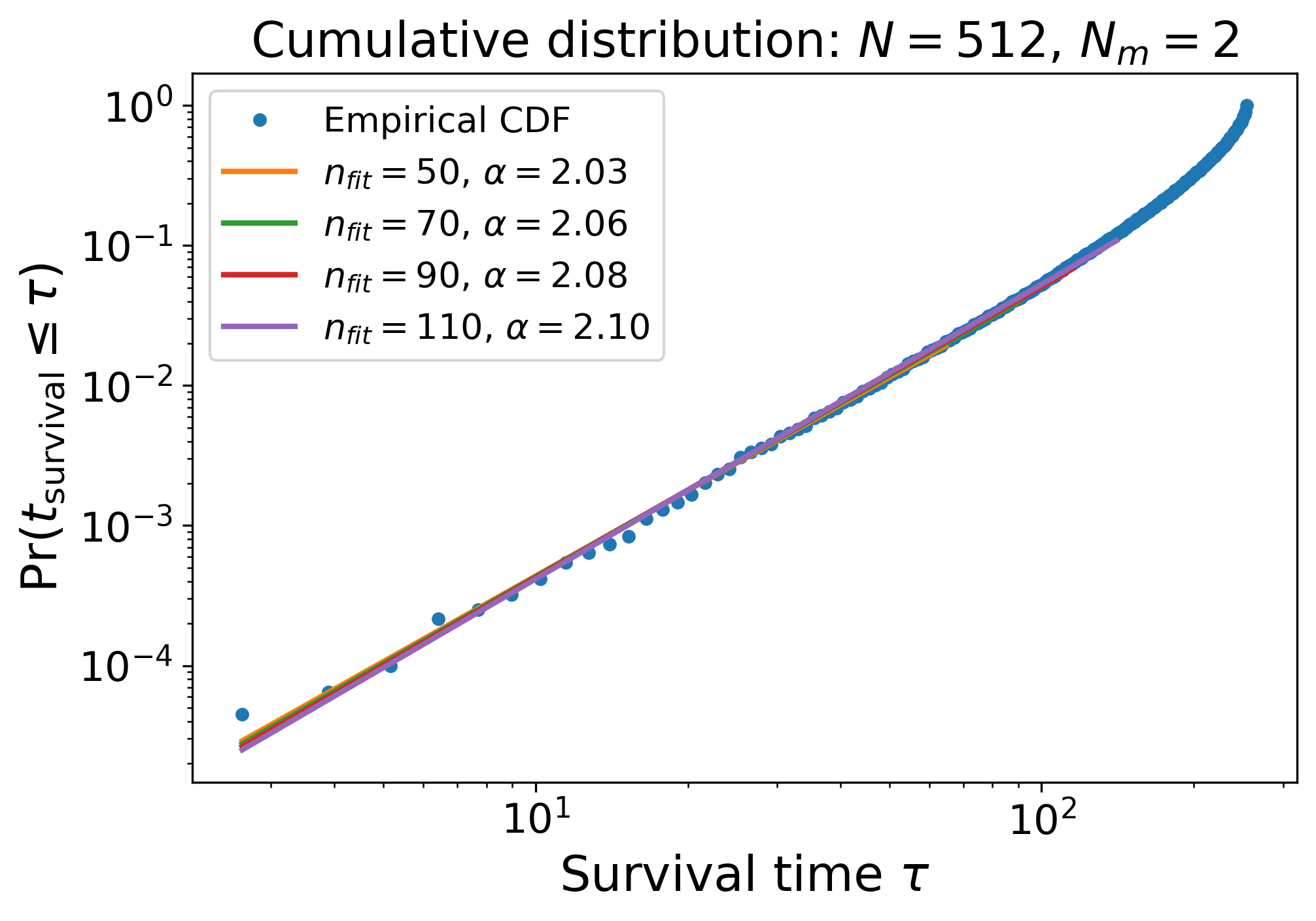}
    \caption{Check of the mean-field behavior of the cumulative survival-time distribution for $N^{(0)}=512$ and $N_m^{(0)}=2$. The power-law fit performed in the early-time regime, up to $\tau_{\max}=50$, gives $\alpha=2.03$, which is very close to the theoretical prediction $\alpha=N_m=2$.}
    \label{fig_mean_field}
\end{figure}

Based on this first analysis, it is also useful to investigate the cumulative
probability that the survival time is shorter than a given time $\tau$. This
quantity is shown in Fig.~\ref{fig_histogram}(b). We observe that the
cumulative probability increases with $\tau$, displaying an initial power-law
regime whose exponent depends on the initial number of mobsters, $N_m^{(0)}$, as
predicted by Eq.~\ref{Eq:power_law_DA}. For $N_m^{(0)}=4$ and $N^{(0)}=256$, the measured exponent is approximately $4.7$ up to $\tau\simeq 100$, corresponding to the early-time regime. Although this value is close to the theoretical prediction,
the ratio $N_m^{(0)}/N^{(0)}$ is not sufficiently small in this case for the asymptotic approximation to be fully accurate. To test the prediction in a more dilute regime, we repeated the experiment for $N^{(0)}=512$ and $N_m^{(0)}=2$. In this case, fitting the cumulative distribution up to $\tau_{\max}=50$ gives
$\alpha\simeq 2.03$, much closer to the expected value $\alpha=N_m^{(0)}=2$, in
agreement with the early-time behavior predicted by Eq.~\ref{Eq:power_law_DA}.
Increasing the fitting range leads to slightly larger exponents:
$\alpha\simeq 2.06$ for $\tau_{\max}=70$, $\alpha\simeq 2.08$ for
$\tau_{\max}=90$, and $\alpha\simeq 2.10$ for $\tau_{\max}=100$. This
systematic increase indicates that deviations from the asymptotic early-time
regime become more visible as larger survival times are included in the fit.
The corresponding log-log fits are shown in Fig.~\ref{fig_mean_field},
confirming the expected mean-field behavior in the dilute regime.
    
This behavior is also consistent with the rank-ordered analysis shown in Fig.~\ref{fig_histogram}(a), reinforcing the conclusion that the persistence of misinformation is strongly affected by the initial number of spreaders.

Finally, we analyze how the initial number of detectives affects the survival-time statistics. For this purpose, we fix $N_{m}^{(0)}=32$ and compute the cumulative probability for different values of $N_{d}^{(0)}$. The results are shown in Fig.~\ref{fig_histogram}(c). The attenuation of the cumulative curves indicates that the extinction of mafia members occurs at significantly shorter times for a greater initial number of detectives. Moreover, the slope, or effective exponent, decreases substantially as $N_{d}^{(0)}$ increases, changing from approximately $110$ for $N_{d}^{(0)}=0$ to about $35$ for $N_{d}^{(0)}=16$. This result shows that the presence of detectives strongly suppresses the long-time survival of misinformation, reducing both its persistence and its ability to remain active over extended periods.

\begin{figure}
    \centering
    \includegraphics[width=1.0\columnwidth]{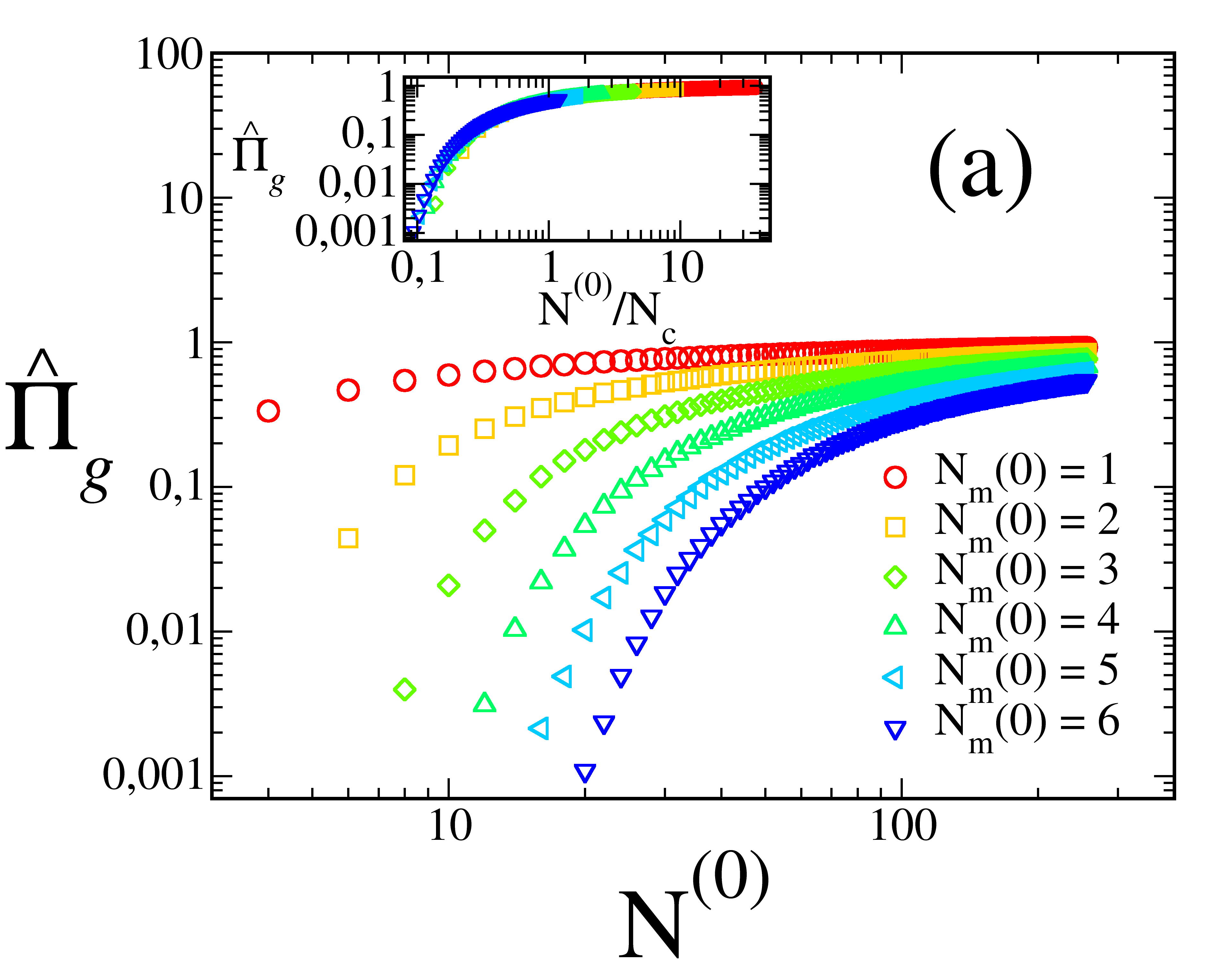}
    \includegraphics[width=1.0\columnwidth]{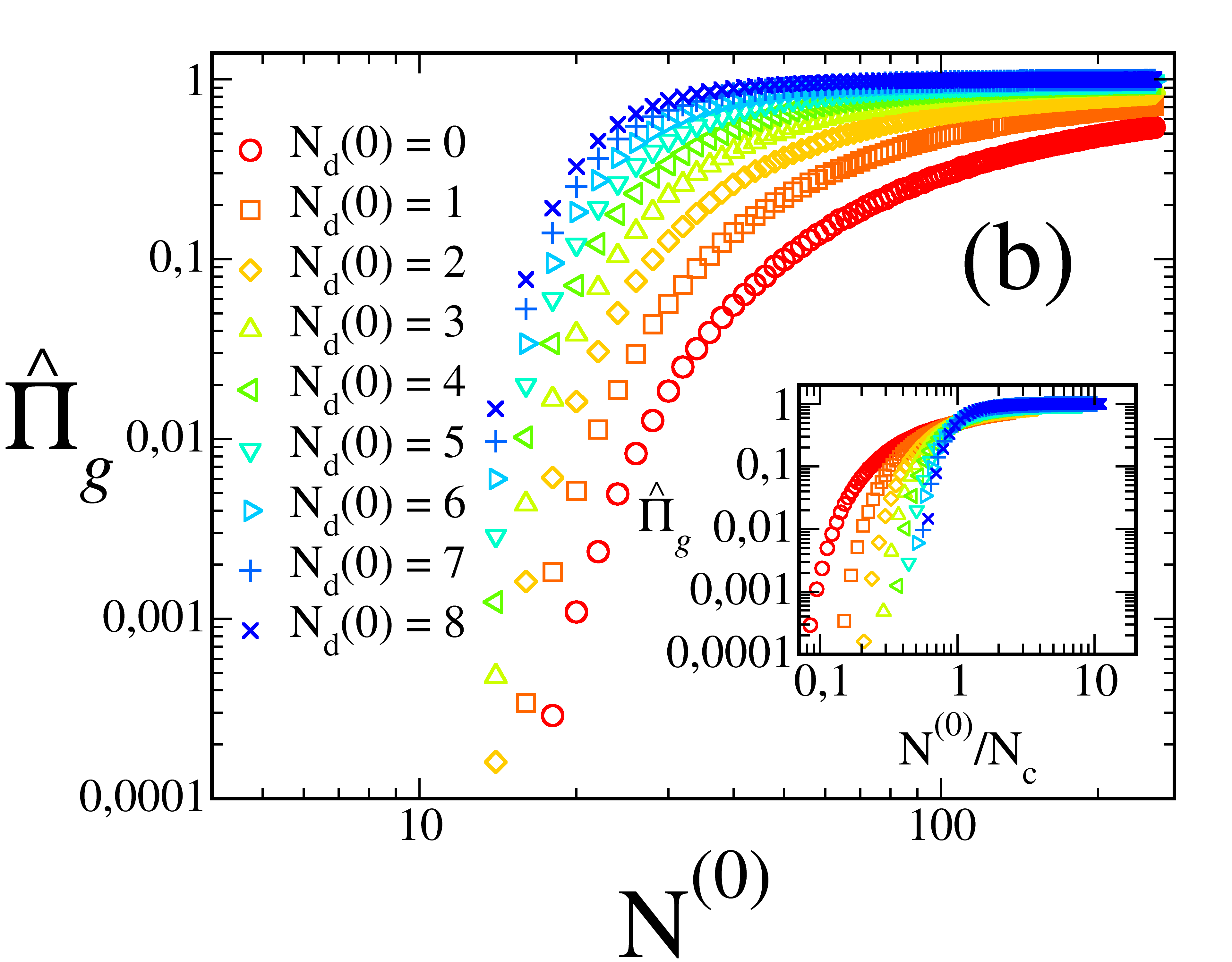}
    \caption{Estimated conditional distribution of villagers winning as a function of the initial population. In plot (a), we show different initial number of werewolves $N_m^{(0)}$ and no detectives. The inset plot shows that, by scaling the initial population with $1/N_c$, the curves collapse, suggesting a universal behavior. In plot (b), we show different values for the number of detectives, while having $N_m^{(0)}=8$. In the inset plot, we observe that the same scaling with $1/N_c$ does not make all curves collapse into a single curve.}
    \label{fig2_scaling}
\end{figure}

Searching for collapse after normalization by a characteristic size follows the finite-size-scaling strategy used in statistical physics to expose common scaling functions across finite systems \cite{FisherBarber1972,Stanley1971}.

Figure \ref{fig2_scaling} illustrates the estimated conditional win probability of the peasants, $\hat{\Pi}_p$, as a function of the initial system size $N^{(0)}$. In Fig. \ref{fig2_scaling}a, we present the results for various initial werewolf populations, $N_m^{(0)}$, in the absence of detectives ($N_d^{(0)} = 0$). Although the curves shift depending on $N_m^{(0)}$, they all exhibit a characteristic sigmoidal profile on a log-log scale. To test for scaling behavior, the inset of Fig. \ref{fig2_scaling}a displays the same probability plotted against the scaled system size $N^{(0)}/N_c$, where $N_c$ represents the critical characteristic population initial size for which $\hat{\Pi}_p = 0.5$. Remarkably, this rescaling leads to an excellent data collapse of all curves onto a master curve. This collapse implies that the initial system size governs the absorption statistics in a universal manner when no detectives are present. To accurately extract the values of $N_c$ for each independent curve where $\hat{\Pi}_p = 0.5$, a cubic spline interpolation scheme was implemented over the discrete simulation data points.

Conversely, Fig. \ref{fig2_scaling}b explores the effect of the initial number of detectives, $N_d^{(0)}$, for a fixed predator density of $N_m^{(0)} = 8$. While the curves retain their qualitative sigmoidal shape, the introduction of detectives breaks the previous scaling symmetry. As shown in the inset of Fig. \ref{fig2_scaling}b, applying the same rescaling factor $N^{(0)}/N_c$ fails to collapse the data. This lack of data collapse indicates that the introduction of a third competing group (detectives) alters the underlying critical behavior, driving the system away from the simple universality class observed in the pure predator-prey scenario.

\section{\label{sec:discussion}Discussion, summaries, and conclusions}

The results obtained through our agent-based formulation of the \textit{Mafia Party} game provide critical insights into how information and group coordination influence the persistence of truth in a society susceptible to malicious manipulation. By mapping the structural properties of the game onto an opinion-dynamics framework, we can interpret our statistical metrics through the lens of sociocompational phenomena.

We have shown analytically and through numerical simulations that the distribution of survival times of mobsters followed a power law form whose exponent is the actual initial size of the mafia, when the total number of players was sufficiently larger than the mafia population. 

When considering the statistics of the game outcome, i.e., late-time dynamics, our results reveal a non-equilibrium phase transition that depends strictly on the composition of the system. In the absence of independent verifiers ($N_d = 0$), there is a perfect data collapse, as showcased in the inset of Fig. \ref{fig2_scaling}, indicating that the system belongs to a unique universality class. Here, the characteristic size $N_c$ acts as a macroscopic threshold: if the population is small ($N^{(0)} \ll N_c$), the microscopic dilution allows coordinated manipulators (mafia) to rapidly bias public consensus. Conversely, sufficiently large system sizes inherently dilute the impact of coordinated fake news, granting the majority a statistical advantage to filter out noise, purely as a consequence of the law of large numbers applied to absorbing trajectories.

The inclusion of detectives ($N_d > 0$) severely breaks this scaling symmetry, as evidenced by the lack of data collapse in Fig. \ref{fig2_scaling}b. From a physical perspective, the verifier subpopulation acts as a relevant field perturbation that alters the phase space. 

\textbf{Funding:} R. da Silva was partially supported by CNPq under grants 304575/2022-4 (Productivity Fellowship – PQ), 406820/2025-2 (Universal), 408596/2024-4 (INCT–NanoQuantSS), and 444867/2024-4 (CNPq/MCTI/FNDCT). E. V. Stock and R. da Silva also acknowledge financial support from FAPERGS (Fundação de Amparo à Pesquisa do Estado do Rio Grande do Sul) under Grant 25/2551-0002529-0.

\bibliographystyle{unsrt}
\bibliography{disinformation_mafia_refs_2}

\end{document}